\documentclass[intlimits,twoside,a4paper]{article}
\usepackage{amsmath,amssymb}
\usepackage{graphicx}
\usepackage{wrapfig}
\usepackage[T2A]{fontenc}
\usepackage[cp1251]{inputenc}

\usepackage{cmpj2}

\issue{2014}{17}{3}{33004}
\doinumber{10.5488/CMP.17.33004}
\title[Universality and self-similar behaviour of non-equilibrium systems]%
{Universality and self-similar behaviour of non-equilibrium systems with non-Fickian diffusion\thanks{In memory of Prof. A.~Olemskoi.}}%

\author[D.O. Kharchenko, V.O. Kharchenko, S.V. Kokhan]{D.O. Kharchenko, V.O. Kharchenko, S.V. Kokhan}

\address{Institute of Applied Physics, Nat. Acad. Sci. of Ukraine,
58 Petropavlivska St., 40000 Sumy, Ukraine}

\date{Received January 28, 2014, in final form May 19, 2014}

\begin{document}

\maketitle

\begin{abstract}
Analytical approaches describing non-Fickian diffusion in complex systems are
presented. The corresponding methods are applied to the study of statistical
properties of pyramidal islands formation with interacting adsorbate at
epitaxial growth. Using the generalized kinetic approach we consider
universality, scaling dynamics and fractal properties of pyramidal islands
growth. In the framework of generalized kinetics, we propose a theoretical model
to examine the numerically obtained data for averaged islands size, the number of
islands and the corresponding universal distribution over the island size.

\keywords complex systems, nonlinear diffusion, pattern formation, fractals
\pacs 05.40.-a, 05.10.Gg, 05.45.Df
\end{abstract}

\section{Introduction}
It is well known that complex physical, chemical and/or biological systems
manifest strong interactions between their elements (atoms, molecules,
individuals/organisms) leading to self-organization processes with anomalous
diffusion, nonequilibrium phase transitions, patterning and the formation of
self-similar objects such as fractals. The cooperative interactions in such
systems result in a reduction of a large number of degrees of freedom binding
the subunits of these systems by means of self-organization into synergetic
entities. These synergetic systems admit low-dimensional description by the
corresponding Fokker-Planck equation (FPE). As was shown previously (see for
example, \cite{TDFrank}), systems manifesting a collective behavior of
interacting species are described in terms of nonlinear FPE (NFPE) capturing
the essential dynamics underlying the observed phenomena with anomalous transport
and generalized (nonextensive) or $Q$-deformed statistics \cite{Tsallis}. The
corresponding nonlinearity in NFPE without a drift component, leads to non-Fickian
diffusion with density/concetration dependent dispersal resulting in anomalous
dynamics as was shown experimentally and theoretically (see, for example,
\cite{Tsallis,KH2010,SOC2003,KK2005,OK2006}).

Nonlinear diffusion equation can be generalized by taking
quasi-chemical reactions into account. In such a case, one arrives at nontrivial scenarios
for nonequilibrium phase transitions \cite{PhysA09}, pattern formation
\cite{PhysD09} and delaying dynamics at phase separation processes
\cite{PhysA2008}. For example, while studying the pattern formation phenomena on surfaces
at deposition from a gaseous phase, one describes the corresponding system by
reaction-diffusion model with field dependent diffusivity. It was shown that
nanosize patterns of adsorbate can emerge due to microscopic interactions of
the deposited particles
\cite{BHKM97,HME98_1,HME98_2,MW2005,M2010,PRE2012,PhysScr2012}. In
\cite{EPL} it was shown that fractal properties of porous-surface
condensates are described by the generalized statistics based on NFPE approach
and by the corresponding theory of multifractals. In the process of pyramidal
islands growth at molecular beam epitaxy, it was found that a structure of
pyramids essentially depends on interactions of the elements forming the pattern
described by a concentration dependent diffusion coefficient
\cite{KP98,LM94,PhysScr2011,EPJB2013}. While studying the arrangement of point defects in
solids at particle irradiation according to the swelling rate theory
\cite{Walgraef1,Walgraef2}, it was found that vacancies can arrange into
nanosize clusters due to their interactions described by a nonlinear diffusion
flux \cite{EPJB2012,UJP2013,CMPh2013,REDS2014}. This effect can lead to
abnormal grain growth dynamics when vacancies segregate on the grain boundaries in a
stochastic manner \cite{PRE2014}. Nonlinear diffusion was
experimentally studied for a large variety of chemical and biological systems
\cite{Murray,Witelski}. In \cite{21,22,26} it was shown that nonlinear
diffusion leads to anomalous dynamics. From this non-complete literature
overview, it follows that in spite of the diversity of this research area, many
phenomena in physical, chemical and biological systems have fundamental
physical mechanisms in common. A study of interacting mechanisms of elements of
complex systems leading to macroscopic self-organization processes with
collective behavior of their species remains an urgent problem in modern
statistical physics during the last two decades.

In this paper we initially present a generalized kinetic approach permitting to
describe the behavior of physical systems with an interaction of their species.
Here, we illustrate a self-similar behavior of the main statistical characteristics
for the systems described by nonlinear diffusion equation. Next, studying
self-organization processes with the surface pattern formation in a system of
interacting adsorbate at molecular beam epitaxy, we use the formalism of
nonlinear kinetic approach to describe the scaling properties of pyramidal islands
growth and the universality of the system behavior.

The work is organized as follows. In section~\ref{s2} using the generalized master
equation we introduce NFPE and discuss the corresponding anomalous dynamics
characterized by nonextensive statistics. Here, we consider the nonlinear
diffusion equation and study the properties of a related solution. In section~\ref{s3} we
discuss the physical reasons responsible for the emergence of nonlinear diffusion flux
in systems with interacting species. Here, (subsection~\ref{s3.1}) we consider a
typical model of reaction-diffusion systems with field dependent diffusivity
related to the surface pattern formation at molecular beam epitaxy. The original
results illustrating the role of particle interactions and scaling dynamics
with universal properties of the system described in the framework of NFPE
approach are collected in subsection~\ref{s3.2}. We conclude in section~\ref{s4}.

\section{Generalized kinetic approach and nonlinear diffusion equation\label{s2}}

The classical Fickian diffusion assumes that there are no interactions between
the random walkers representing individuals in a system. If these walkers
interact, the diffusivity or the mobility can change in the presence of walkers
and become explicitly dependent functions of the local concentrations of the moving
particles. This leads to NFPE as a generalization of ordinary linear FPE where
transition probabilities depend on occupation probability densities of initial
and arrival states. Following \cite{Kan2001}, NFPE can be constructed from
the generalized master equation written for a probability density function
(pdf) $p(\mathbf{r},t)$ to find a particle in the state $\mathbf{r}$ at the
time $t$. This equation has the form
\begin{equation}\label{me}
\frac{\partial p(\mathbf{r},t)}{\partial t}=\int \left[ w(\mathbf{r}',
\mathbf{r}-\mathbf{r}')\gamma(p',p)-w(\mathbf{r},
\mathbf{r}'-\mathbf{r})\gamma(p,p')\right]{\rm d}\mathbf{r}',
\end{equation}
where $w(\mathbf{r},\mathbf{r}')$ is the transition rate between two states
located at $\mathbf{r}$ and at $\mathbf{r}'$ which depends on the nature of
interactions between the particle and the bath; it is a function of starting
$\mathbf{r}$ and arrival $\mathbf{r}'$ sites. The factor $\gamma(p,p')$ is an
arbitrary function of the particle populations of both the initial and the
arrival sites. This function satisfies the following conditions: $\gamma(0,p')=0$
meaning that if the initial site is empty, the transition probability is equal
to zero; $\gamma(p,0)\ne 0$ requires that if the arrival site is empty,
the transition probability should depend on the population of the initial site. In the
case $\gamma(p',p)=p'$ and $\gamma(p,p')=p$, we have the standard liner kinetics
when equation~(\ref{me}) reduces to the Chapman--Kolmogorov equation.

In the diffusion limit one can expand slowly varying functions $\gamma$ in
equation~(\ref{me}) and obtain NFPE in the form \cite{Kan2001}
\begin{equation}\label{fpe}
\frac{\partial p}{\partial t}=-\nabla \left[f(\mathbf{r})-\nabla
D(\mathbf{r})\right]\gamma(p)+\nabla \left[D(\mathbf{r})\gamma(p)\frac{\partial
\ln\kappa(p)}{\partial p}\nabla p\right],
\end{equation}
where $\nabla\equiv\partial/\partial \mathbf{r}$. The drift and diffusion terms
are: $f(\mathbf{r})\equiv\int
\rd\mathbf{r}'\mathbf{r}'w(\mathbf{r},\mathbf{r}')$, $2D(\mathbf{r})\equiv\int
\rd\mathbf{r}' {\mathbf{r}'}^2w(\mathbf{r},\mathbf{r}')$. Here,
$\gamma(p)=\gamma(p,p)$, and the function $\kappa(p)>0$ is defined through the
condition
\begin{equation}
\frac{\partial \ln\kappa(p)}{\partial p}=\left[\frac{\partial }{\partial
p}\ln\frac{\gamma(p,p')}{\gamma(p',p)}\right]_{p=p'}.
\end{equation}
Formally, the functions $\gamma(p)$ and $\kappa(p)$ can be represented through
densities of the initial $a(p)$ and arrival $b(p)$ states as follows:
$\gamma(p)=a(p)b(p)$, $\kappa(p)=a(p)/b(p)$ \cite{Chavanis}.

In stationary equilibrium state, the solution $p_{\mathrm{s}}=p(\mathbf{r},\infty)$ of
equation~(\ref{fpe}) takes the form
\begin{equation}\label{potentialU}
\ln\kappa({p_{\mathrm{s}}})= U_{0}-U_{\mathrm{ef}}(\mathbf{r}), \qquad
U_{\mathrm{ef}}(\mathbf{r})=-\int\limits^\mathbf{r}\frac{f(\mathbf{r}')}{D(\mathbf{r}')}{\rm
d}\mathbf{r}'+\ln D(\mathbf{r}),
\end{equation}
where $\gamma(p)$ remains an arbitrary function; $U_0$ takes care of
normalization condition $\int p_{\mathrm{s}}{\rm d}\mathbf{r}=1$. A nonlinearity of the
Fokker-Planck equation is defined through the form of the function $\kappa(p)$:
at $\ln \kappa(p)=\ln p$, we move to the Boltzmann-Gibbs statistics; a
$Q$-deformation of the logarithm $\ln\kappa\to\ln_{Q}\kappa$ promotes Tsallis
statistics \cite{Tsallis}, where $\ln_Q\kappa=(\kappa^{1-Q}-1)/(1-Q)$.

In the simplest case with $f(\mathbf{r})=0$ and $D=\mathrm{const}$, one can consider the time
dependent solution of the nonlinear diffusion equation as reduced NFPE. Here, by
taking $\gamma=\kappa=p$ one gets
\begin{equation}\label{ndif}
\frac{\partial p}{\partial t}=\nabla \left[\mathcal{D}(p)\nabla p\right],\qquad
\mathcal{D}(p)\equiv D\gamma(p)\kappa(p)^{-Q}\frac{{\rm d}\kappa(p)}{{\rm
d}p}=Dp^{1-Q}.
\end{equation}
As was shown in \cite{Murray,Ts96,St97}, this equation has an exact solution
in one dimension at $|r|\leqslant r_0 (t/t_0)^{H}$:
\begin{equation}
p(r,t)=\left(\frac{t_0}{t}\right)^{H}\left[1-\left(\frac{rt_0^{H}}{r_0t^{H}}\right)^2\right]^{\frac1{1-Q}},\qquad
H=\frac{1}{3-Q}\,, \qquad r_0=\frac{N_0\Gamma\left(\frac{1}{1-Q}+\frac
32\right)}{\pi^{1/2}\Gamma\left(\frac{2-Q}{1-Q}\right)}\,, \qquad
t_0=\frac{r_0^2(1-Q)}{2D(3-Q)}\,,
\end{equation}
where $N_0$ is the initial number of particles at the origin. At $|r|>r_0
(t/t_0)^{H}$, one has $p(r,t)=0$. It follows that such a nonlinear diffusion
equation admits scaling $\langle(r-\langle r\rangle)^2\rangle\propto t^{2H}$,
whereas ordinary Brownian diffusion corresponds to $H=1/2$ with $Q=1$. Scaling
regimes in a self-affine phase space characterized by $D(r,p)=p^{1-Q}r^\Delta$
were studied in \cite{KK2005}. An automodel solution of the corresponding
nonlinear diffusion equation is characterized by the Hurst exponent $H$ in the
form $H=1/(3-Q-\Delta)$, whereas the related pdf is well-described by the
Tsallis form. Therefore, anomalous dynamics can be controlled either by Tsallis
parameter $Q$ or by the exponent $\Delta$.

In the case of the constant drift term [$f(\mathbf{r})=\mathrm{const}$], the
corresponding simulations of the time series, satisfying Tsallis statistics,
have shown clusterization of time series manifesting a self-similar regime
characterized by fractal properties \cite{Feder,OK2006}. A study of the
stochastic dynamics corresponding to NFPE with a power law dependence
$\gamma(p)$ was done in \cite{LBorland}; here, anomalous dynamics was
considered using correlation analysis. The relation between NFPE and Levy-type
diffusion was discussed in \cite{Ts96,Ts95}. The application of the
generalized statistics followed by NFPE with anomalous diffusion emergent at
self-organized criticality regimes was studied in \cite{SOC2003}, where
scaling laws were discussed for avalanche sizes dynamics and the corresponding power-law
distributions.

Therefore, the nonlinear kinetic approach is applicable to a description of a
universality and scaling behavior of complex systems characterized by field
dependent diffusivity related to interactions between system elements. In the
next section we study the pattern formation processes by considering the reaction
diffusion model that describes epitaxial growth with interacting adsorbate. We
use the nonlinear kinetic formalism to analytically describe the scaling regimes
and universality of probability density functions of pyramidal islands over
their sizes.

\section{Universality and scaling regimes of pattern formation at epitaxy\label{s3}}

In the previous section we have considered the nonlinear diffusion equation, where
density dependent diffusion coefficient was defined through occupation
probabilities of the initial $a(p)$ and arrival $b(p)$ states. In this section, while
 studying the pyramidal islands growth at epitaxy, we illustrate that the
corresponding nonlinear diffusivity can be immediately obtained considering the
 interactions of adsorbed particles. Universality and scaling regimes of
the arranged pyramidal islands will be studied with the help of NFPE presented in the
previous section.

\subsection{Phase field model of pyramidal islands growth\label{s3.1}}
Considering a system of interacting adsorbate with free diffusion on a surface
and the motion of mobile particles caused by their interactions, one can directly obtain the
diffusion flux with a field dependent diffusion coefficient. In such a
case, instead of the probability density $p$, the relevant quantity that can be
used is a coverage field $\rho$ (concentration of atoms/moleculas adsorbed by
the surface). An evolution of this quantity is described by a
reaction-diffusion equation of the form $\partial_t
\rho=R(\rho)-\nabla\cdot\mathbf{J}$, where $R(\rho)$ is the reaction term
including adsorption/desorption and/or additional nonequilibrium processes
related to the formation of islands, or oxides; $\mathbf{J}$ is the diffusion flux.

At epitaxial growth, the reaction term $R(\rho)$ is defined through a deposition
rate characterized by the flux $F_0$ of arriving particles on a surface and
by desorption reaction $-\rho/\tau$, where $\tau$ is the relaxation time. Next, we
assume that desorbed particles have lateral interactions described by the
attractive potential $U(\mathbf{r})$. This leads to renormalization of the
desorption rate $\tau^{-1}=\tau_{0}^{-1}\exp(U/T)$, where $\tau_{0}^{-1}$ is
the constant related to a hopping rate, $T$ is the temperature measured in
energetic units.

In general case, the diffusion flux can be written in the standard form
$\mathbf{J}=-\mathcal{D}(\rho)\nabla\frac{\delta \mathcal{F}}{\delta \rho}$,
where $\mathcal{D}(\rho)$ is a field dependent diffusivity and
$\mathcal{F}[\rho]$ is a free energy. Let us define the related diffusivity and
the free energy. For the free diffusion of mobile species, one has a standard
definition $\mathbf{J}_{\mathrm{dif}}=-D\nabla \rho$, with $D=\mathrm{const}$. A strong local
bond induced by the interaction of adsorbed particles can be described by the
potential $U[\rho(\mathbf{r})]=-\int
u(\mathbf{r}-\mathbf{r}')\rho(\mathbf{r}'){\rm d}\mathbf{r}'$, where
 $u(r)$ is a spherically symmetric function depending on the nature of
the system. Therefore, the corresponding thermodynamical (chemical) force
$\mathbf{f}=-\nabla(U/T)$ induces speed $\mathbf{v}=D\mathbf{f}$ of mobile
species (Einstein's relation). Hence, the associated flux $\mathbf{v}\rho$ is
possible to $(1-\rho)$ free sites. In such a case, for this flux one has
$\mathbf{J}_{o}=-D\rho(1-\rho)\nabla (U/T)$. Therefore, the total flux
$\mathbf{J}=\mathbf{J}_{\mathrm{dif}}+\mathbf{J}_o$ can be written as
\begin{equation}\label{Jtot}
\mathbf{J}=-D\left[\nabla \rho+\rho(1-\rho)\nabla (U/T)\right].
\end{equation}
Assuming that an interaction length $r_0$ between two interacting species is
small compared to a diffusion length $\ell=\sqrt{D\tau_{0}}$, i.e., $r_0\ll
\ell$, one can write $U(\rho(r))\simeq-\epsilon \rho(r)$ with $\epsilon=\int
u(r){\rm d}r$. The presented formalism is general and can be applied to a large
class of systems: semiconductors and metals, where the quantity $\epsilon$
depends on concrete material properties. Using these assumptions one can
rewrite the total diffusion flux (\ref{Jtot}) through the free energy
functional $\mathcal{F}[\rho]$ in the standard form
\begin{equation}\label{Jtott}
\mathbf{J}=-\mathcal{D}(\rho)\nabla\frac{\delta\mathcal{F}}{\delta \rho}; \qquad
\mathcal{D}(\rho)=\frac{D}{T}\rho(1-\rho), \qquad \mathcal{F}=\int{\rm
d}\mathbf{r} \left\{T\left[\rho\ln
\rho+(1-\rho)\ln(1-\rho)\right]-\frac{\epsilon}{2}\rho^2\right\}.
\end{equation}
Therefore, the reaction diffusion model of a system with an interacting
adsorbate can be written in the form $\partial_t\rho=F-\rho \re^{-\varepsilon
\rho/\theta}-\nabla\cdot \mathbf{J}$, where dimensionless quantities
$t'=t/\tau_{0}$, $\mathbf{r}'=\mathbf{r}/\ell$, $F=F_0\tau_{0}$,
$\theta=T/T_0$, $\varepsilon=\epsilon/T_0$ are used; $T_0$ is the bath
temperature. Next, we drop the primes for convenience. This approach is widely used
to study the pattern formation processes in chemical systems, at condensation under
deterministic (see \cite{BHKM97,HME98_1,HME98_2}) and stochastic
conditions (see \cite{MW2005,M2010,PRE2012,PhysScr2012}), at epitaxial
growth \cite{PhysScr2011,EPJB2013}, at the formation of point defect clusters due
to irradiation effect \cite{EPJB2012,UJP2013,CMPh2013,mirz1} and at other
physical systems manifesting interactions between their elements.

The obtained model cannot be used immediately to model the pyramidal islands
formation because it does not take into account discrete steps (sharp
interfaces) between terraces in pyramids. The problem of sharp interface
modelling at the pyramidal islands formation can be solved using the phase field
approach proposed by Liu and Metiu \cite{LM94} and Karma and Plapp \cite{KP98}.
The idea of the phase field approach lies in introducting an order parameter
$\phi(\mathbf{r},t)$ that indicates the phase at a particular position. In our case,
the phase field $\phi$ describes the surface height in units of monoatomar
layers. According to \cite{LM94}, local stable minima of the order
parameter relate to terraces whereas rapid spatial variation of $\phi$
corresponds to the positions of steps. In the framework of the phase field approach
\cite{KP98}, one has
$
\tau_\phi\partial_t \phi=-\frac{\delta \mathcal{H}}{\delta \phi}$. Here,
$\tau_\phi$ is the characteristic scale of the time of attachment of adatoms at
the step. $\mathcal{H}$ is the effective Hamiltonian, $\mathcal{H}=\int{\rm
d}{\mathbf{r}}[\varpi^2(\nabla\phi)^2/2+g(\phi,x)]$, with the density
$g(\phi,x)=(2\pi)^{-1}\cos(2\pi[\phi-\phi_{\mathrm{s}}])-\lambda
x[\phi+(2\pi)^{-1}\sin(2\pi[\phi-\phi_{\mathrm{s}}])]$. Here, $\varpi$ stands for the width
of the step, $\lambda$ is the dimensionless coupling constant, $\phi_{\mathrm{s}}/2$ is
the height of the initial substrate. This model reduces to the solid-on-solid
model when the coupling between two fields is a constant supersaturation. The
term incorporating $1+\cos(\pi[\phi-\phi_{\mathrm{s}}])$ admits that minima of the free
energy $\mathcal{H}$ are possible at $\phi-\phi_{\mathrm{s}}=2n+1$, independently of the
adatom concentration \cite{KP98}. According to the phase field approach, the
corresponding dynamics of the coverage field is described by the equation $
\partial_t \rho=R(\rho)-\nabla\cdot
\mathbf{J} -\frac{1}{2}\partial_t \phi$ \cite{KP98}. This model is relevant to the case of a constant adsorbate temperature.

It is known that the temperature of the growing surface can be changed locally
at adsorption/de\-sorption processes: when atom becomes adatom, the temperature
increases locally, it decreases when the desorption of adatom occurs. Moreover,
the temperature can be increased due to the effect of the source of atoms
described by $F$. Using the above mechanisms for the temperature variations, one can
write the evolution equation for the temperature field in the dimensionless form
\cite{EPJB2013}
\begin{equation}\label{9}
\mu{\partial_t \theta}=1-\theta+\chi \Delta \theta+r_a F\rho+r_b{\partial_t
   \rho}.
\end{equation}
The relaxation of the temperature to the bath temperature is described by two
first terms in the right-hand side with a relaxation time $\tau_T$, where
$\mu=\tau_T/\tau_0$; $\chi$ plays the role of thermal diffusivity. The fourth
term describes re-heat of the surface with an intensity $r_a$ due to energy
exchange with the environment by deposition flux $F$. This is a standard
assumption and it is widely used considering temperature instabilities in
chemical reactions (see, for example \cite{Horst,mirz1}). Such temperature
instability can be caused by reorganization of the surface due to annihilation
of defects on the surface and their motion to sinks. The fourth term in
equation~(\ref{9}) does not directly take into account a change of the temperature for curved
steps. Such temperature change can be described by the phase field. In
our consideration, this effect is effectively taken into account through the
introduction of the last term in equation~(\ref{9}). It also relates to the local heating
($\partial_t \rho>0$) or cooling ($\partial_t \rho<0$) during
adsorption/desorption processes with intensity $r_b$; generally, it is
responsible for the formation of a curved step.

The total system of three equations describing epitaxial growth generalized by
introducing fluctuation terms is \cite{EPJB2013}
\begin{equation}
\begin{cases}
	\partial_t \rho=&F-\rho \re^{-\varepsilon \rho/\theta}-\nabla\cdot \mathbf{J}
	-\frac{1}{2}\partial_t \phi+\zeta_\rho(\mathbf{r},t),\\
	\vartheta \partial_t \phi=&\varpi^2\Delta\phi-\partial_\phi g(\rho,\phi)+\zeta_\phi(\mathbf{r},t),\\
   \mu{\partial_t \theta}=&1-\theta+\chi \Delta \theta+r_a F\rho+r_b{\partial_t
   \rho}+\zeta_\theta(\mathbf{r},t),
 \label{EQ3}
 \end{cases}
\end{equation}
where $\vartheta=\tau_\phi/\tau_0$. The last terms in equation~(\ref{EQ3}) are stochastic
sources responsible for statistical description of the system dynamics:
$\langle\zeta_i\rangle=0$,
$\langle\zeta_i(\mathbf{r},t)\zeta_j(\mathbf{r}',t')\rangle=\delta_{i,j}\delta(t-t')\delta(\mathbf{r}-\mathbf{r}')$,
where $i,j\in\{\rho,\phi,\theta\}$.

\begin{figure}[!h]
\centering
\includegraphics[width=70mm]{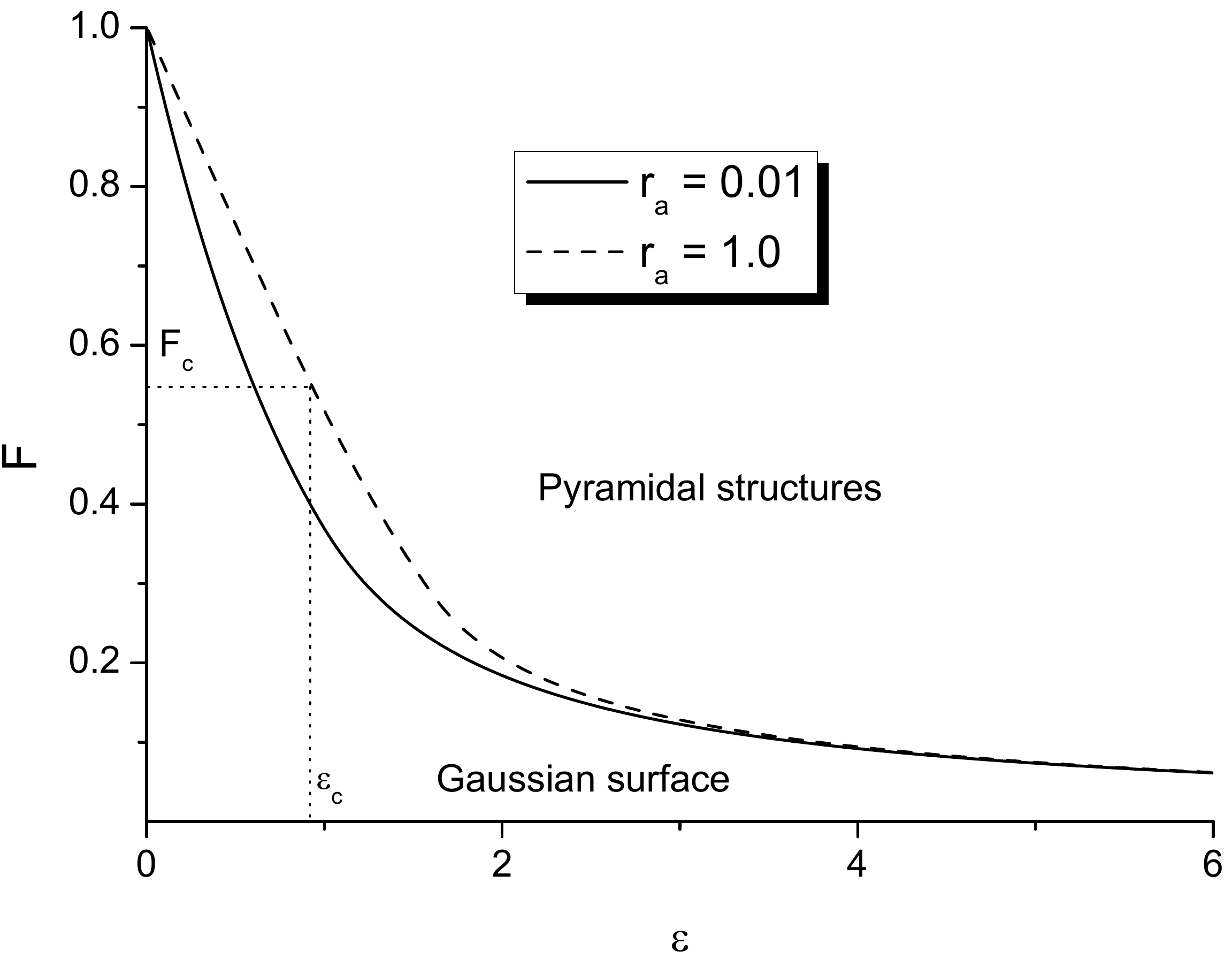}
\caption{Phase diagram for pyramidal structures formation at $\mu=10$,
$\chi=10$, $\varpi=2$, $\lambda=10$, $r_b=0.05$, $\vartheta=1$\label{fig0}.}
\end{figure}
\begin{figure}[!h]
\centering
\includegraphics[width=120mm]{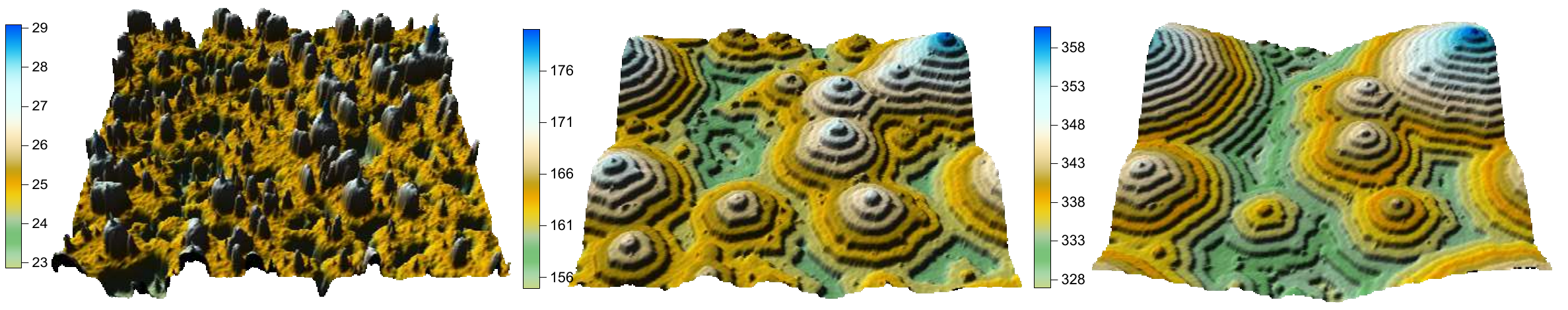}
\caption{(Color online) Evolution of the phase field at $t=5$, 30, and 60
(first second and third column, respectively). Other parameters are:
$\varepsilon=4$, $F=4$, $\mu=10$, $\chi=10$, $\varpi=2$, $\lambda=10$,
$r_a=0.01$, $r_b=0.05, \vartheta=1$\label{evol}.}
\end{figure}

Using the linear stability analysis, one can find critical values of main control
parameters $F$ and $\varepsilon$ bounding a domain of their values
corresponding to the formation of pyramidal islands (see figure~\ref{fig0}). At
$\varepsilon<\varepsilon_{\mathrm{c}}$ and $F<F_{\mathrm{c}}$, the averaged phase field $\langle
\phi\rangle$ does not grow with time meaning the formation of a surface with
Gaussian profile related to the effect of fluctuation terms in equation~(\ref{EQ3}).
In the opposite case, one has $\partial_t\langle \phi\rangle>0$, resulting
 in an increase of the surface height and in the formation of well organized pyramidal
structures on the growing surface. In \cite{PhysScr2011,EPJB2013}, it was
shown that the morphology of pyramids essentially depends on the effect of the
nonlinear diffusion term in the flux of the adsorbate. Snapshots of a typical
evolution of the system are shown in figure~\ref{evol} at $F>F_{\mathrm{c}}$. In our
simulations, we use a square lattice with linear size $L=256$ sites and periodic
boundary conditions. We choose a time step $\Delta t=2.5\times 10^{-4}$ and the
mesh size $l=1$. The Gaussian distributed field
$\phi(\mathbf{r},0)$ with $\langle \phi(\mathbf{r},0)\rangle=0$ and $\langle
[\delta \phi(\mathbf{r},0)]^2\rangle=0.1$ was taken as initial conditions. For initial coverage
and temperature fields we use $\rho(\mathbf{r},0)=0$ and
$\theta(\mathbf{r},0)=1$.
It is seen that at the initial stages, small islands of adsorbate emerge. Here,
the temperature field can be locally changed. During the system evolution, such
islands become centers of pyramidal patterns, and the corresponding pyramids
connect each other by terraces of equivalent heights (see figure~\ref{evol}).

\subsection{Results and discussions\label{s3.2}}

\begin{figure}[ht]
\centerline{\includegraphics[width=0.56\textwidth]{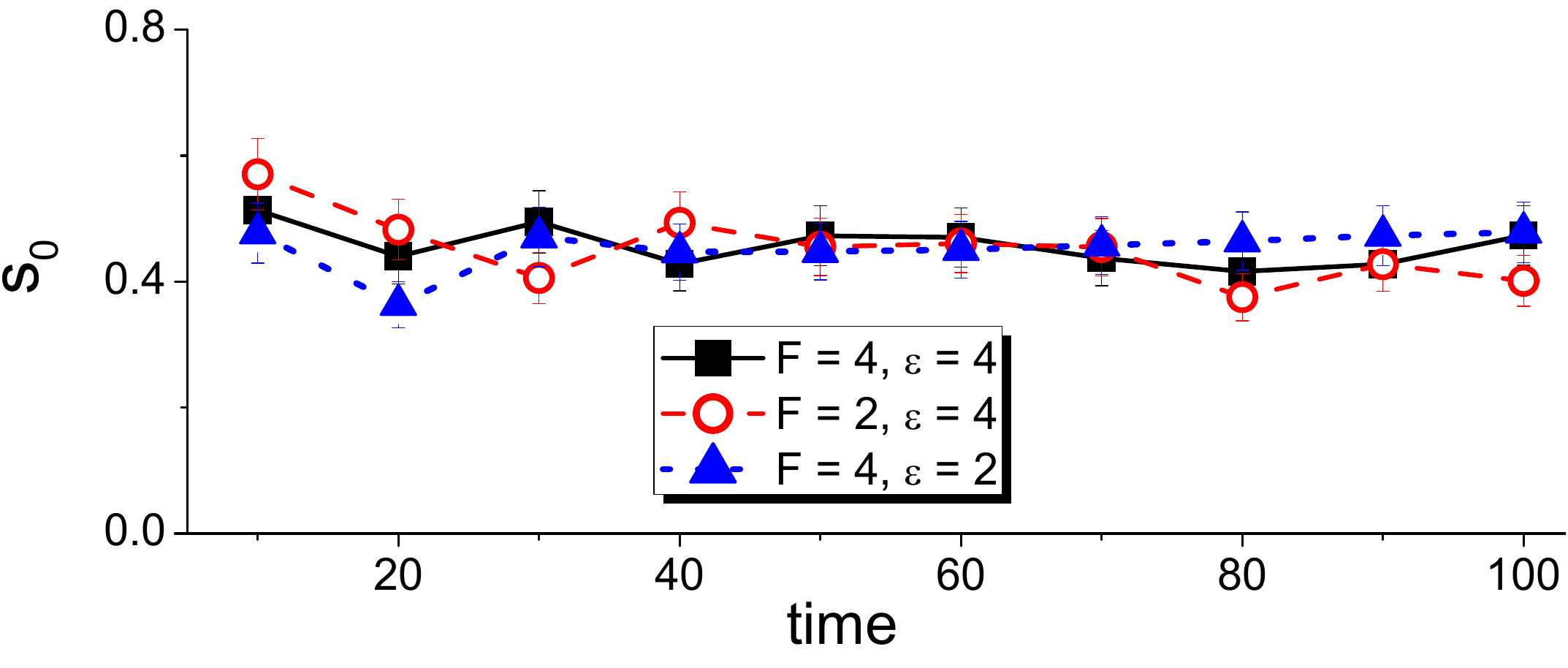}}
\centerline{(a)}%
\includegraphics[width=0.47\textwidth]{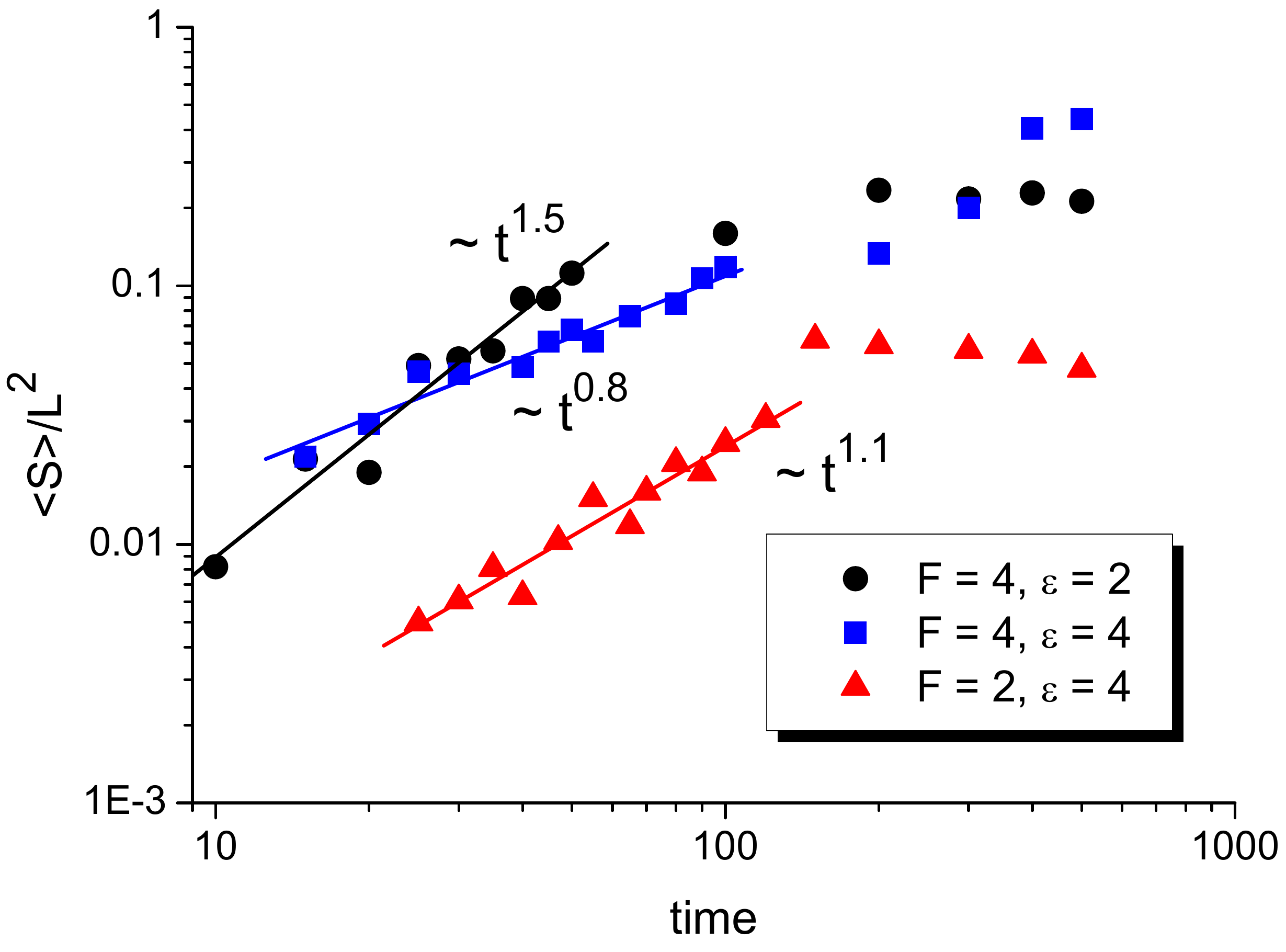}%
\hfill%
\includegraphics[width=0.47\textwidth]{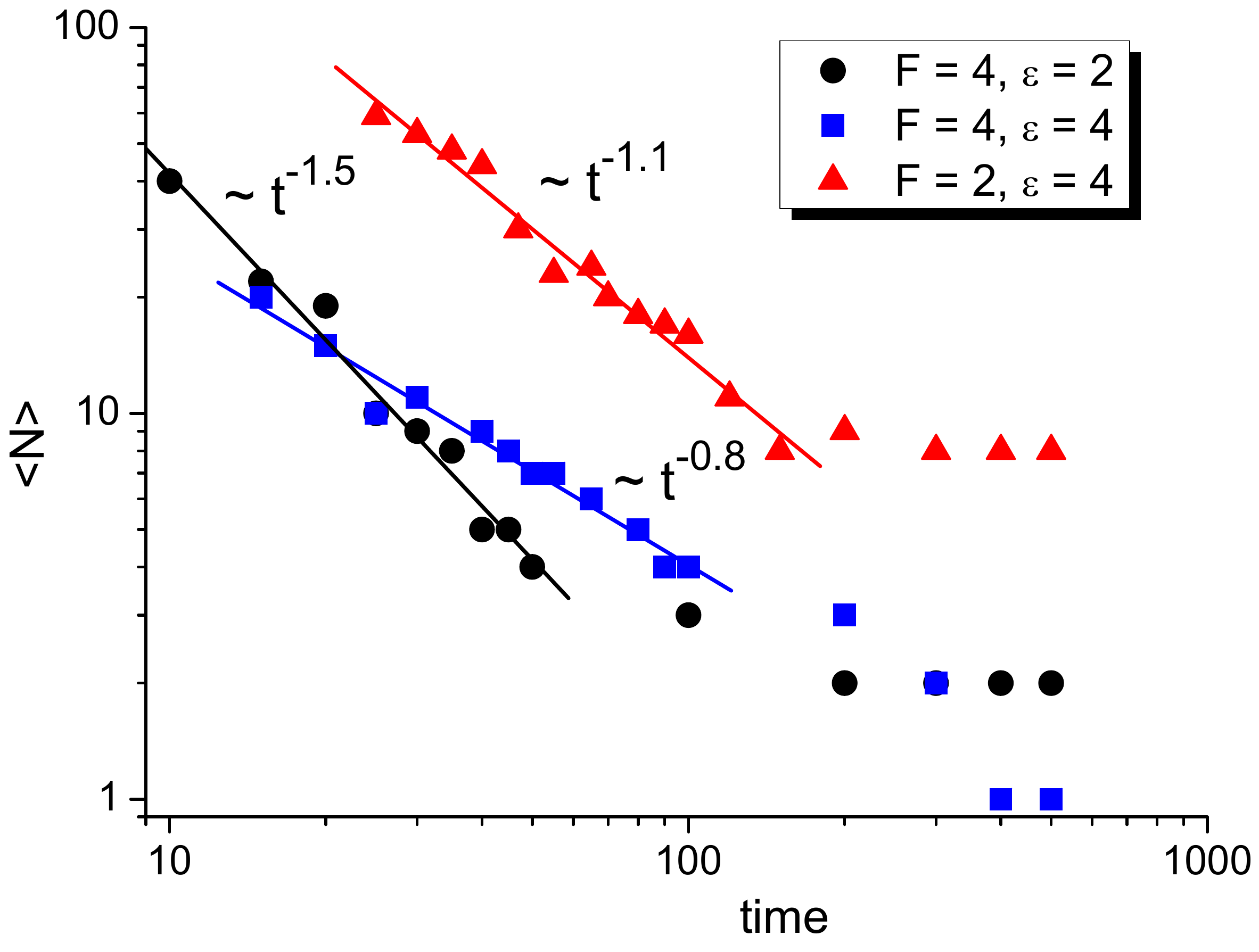}%
\\
\parbox[t]{0.48\textwidth}{%
\centerline{(b)}%
}%
\hfill%
\parbox[t]{0.48\textwidth}{%
\centerline{(c)}%
}%
\caption{(Color online) Time dependencies of the total islands area $S_0$(a),  averaged
island area (b) and the number of islands (c) at a different set of the system
parameters. Other parameters are the same as in figure~\ref{evol}\label{NS}.}
\end{figure}

Let us study the universal dynamics of the islands growth. To this end, we compute
the average islands area $\langle s(t)\rangle$ at a half-height of the whole
system of pyramids emergent at different times and compute the number of the
corresponding islands $N(t)$. It was found that there is a strong relation
between $\langle s(t)\rangle$ and $N(t)$: $\langle s(t)\rangle\propto
[N(t)]^{-1}$, where $\langle s(t)\rangle\propto t^\beta$, $\beta$ is the
scaling exponent.
This relation means that the total area of all islands $S_0$
calculated at different times should remain constant [see figure~\ref{NS}~(a)].
In other words, pdf of the island area distribution $\mathcal{N}(s,t)$ should
satisfy the following two criteria: $\int_0^\infty \mathcal{N}(s,t){\rm
d}s=N(t)$ and $\int_0^\infty s \mathcal{N}(s,t){\rm d}s=S_0\equiv \mathrm{const}$. The
first one defines the number of islands, whereas the second one corresponds to the
surface conservation law. This law is applicable only at the first stages of the
pyramids growth, whereas at late stages one gets only one large constantly growing pyramid. The corresponding dependencies of $\langle s(t)\rangle$ and $N(t)$
are shown in figures~\ref{NS}~(b), (c).

It is seen that the scaling exponent $\beta$ depends on the main system
parameters reduced to $F$ and $\varepsilon$, and takes the values smaller or larger
than 1 related to the normal growth regime: $\langle s(t)\rangle\propto t$.

\begin{figure}[!h]
\includegraphics[width=0.47\textwidth]{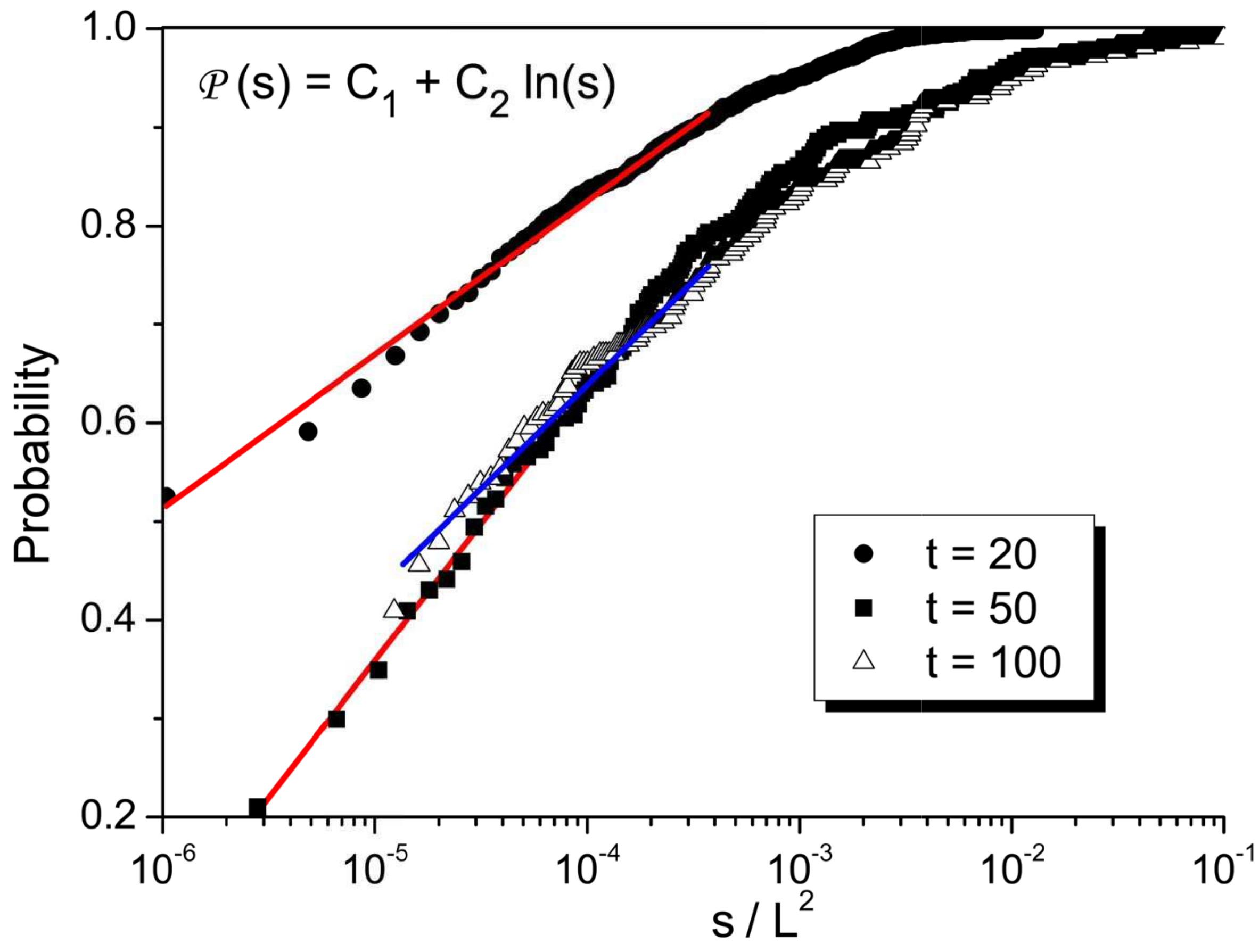}%
\hfill%
\includegraphics[width=0.47\textwidth]{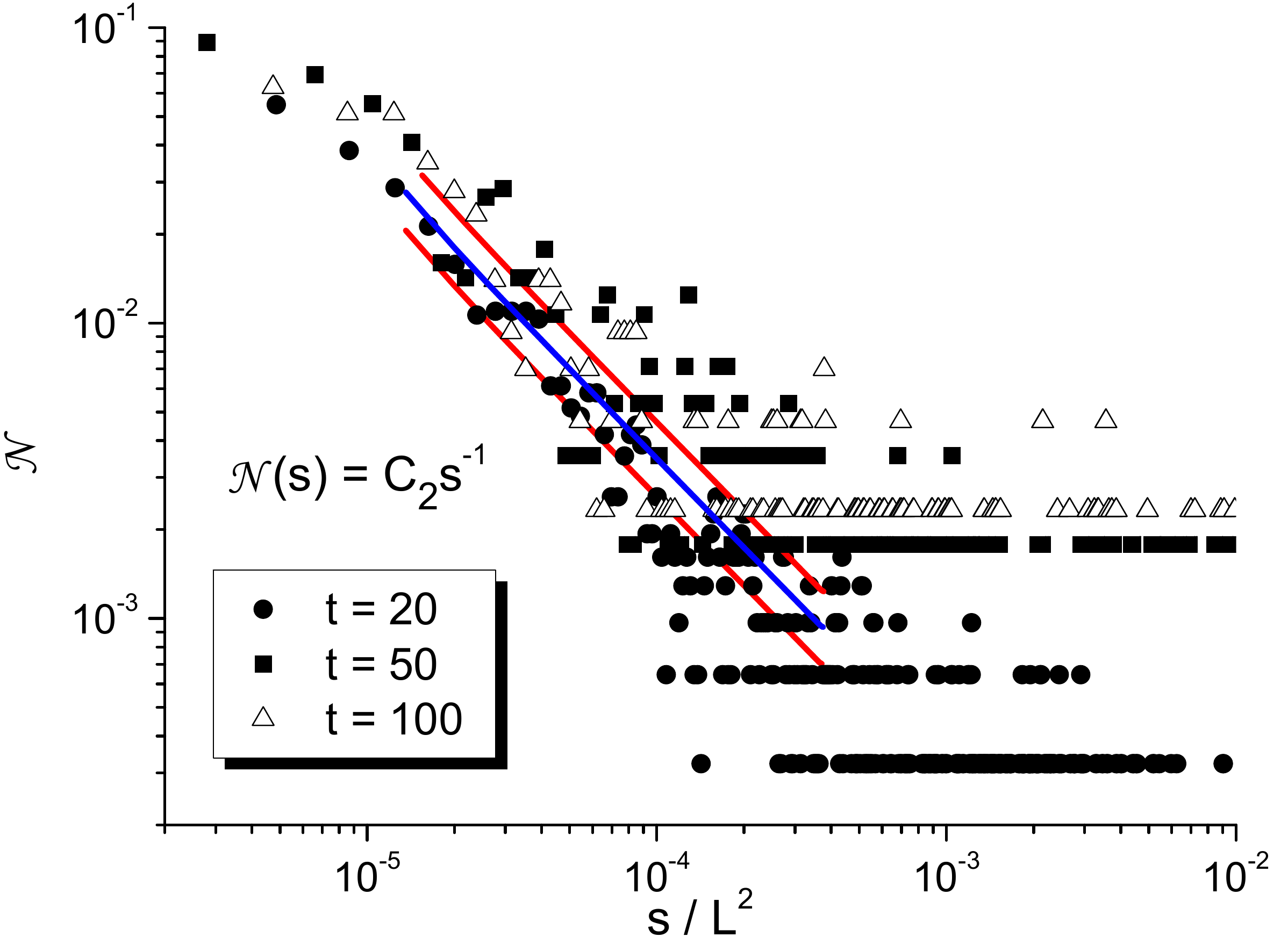}%
\\%
\parbox[t]{0.48\textwidth}{%
\centerline{(a)}%
}%
\hfill%
\parbox[t]{0.48\textwidth}{%
\centerline{(b)}%
}%
\caption{(Color online) Integral distribution function (a) and differential distribution
function (b) at different times at $F=4$, $\varepsilon=4$ (insertion is a
typical snapshot of pyramidal islands at half height). Other parameters are the
same as in figure~\ref{evol}\label{pdfs}.}
\end{figure}
To characterize the distribution of the islands area, we compute the integral
distribution function $\mathcal{P}(s,t)$ at first and get the corresponding pdf
$\mathcal{N}={\rm d}\mathcal{P}/{\rm d}s$ using numerical differentiation (see
figure~\ref{pdfs}).
From the obtained dependencies it follows that the universal behavior of both
$\mathcal{P}(s,t)$ and $\mathcal{N}(s,t)$ at different times is realized.
Moreover, the corresponding fitting procedure gives logarithmic approximation
of $\mathcal{P}(s)$ and algebraic dependencies in the form of the Zipf law
$\mathcal{N}(s)=C_2 s^{-1}$ at different times, where the normalization
parameter is a function of time, i.e., $C_2=C_2(t)$.

\begin{figure}[!t]
\includegraphics[width=0.46\textwidth]{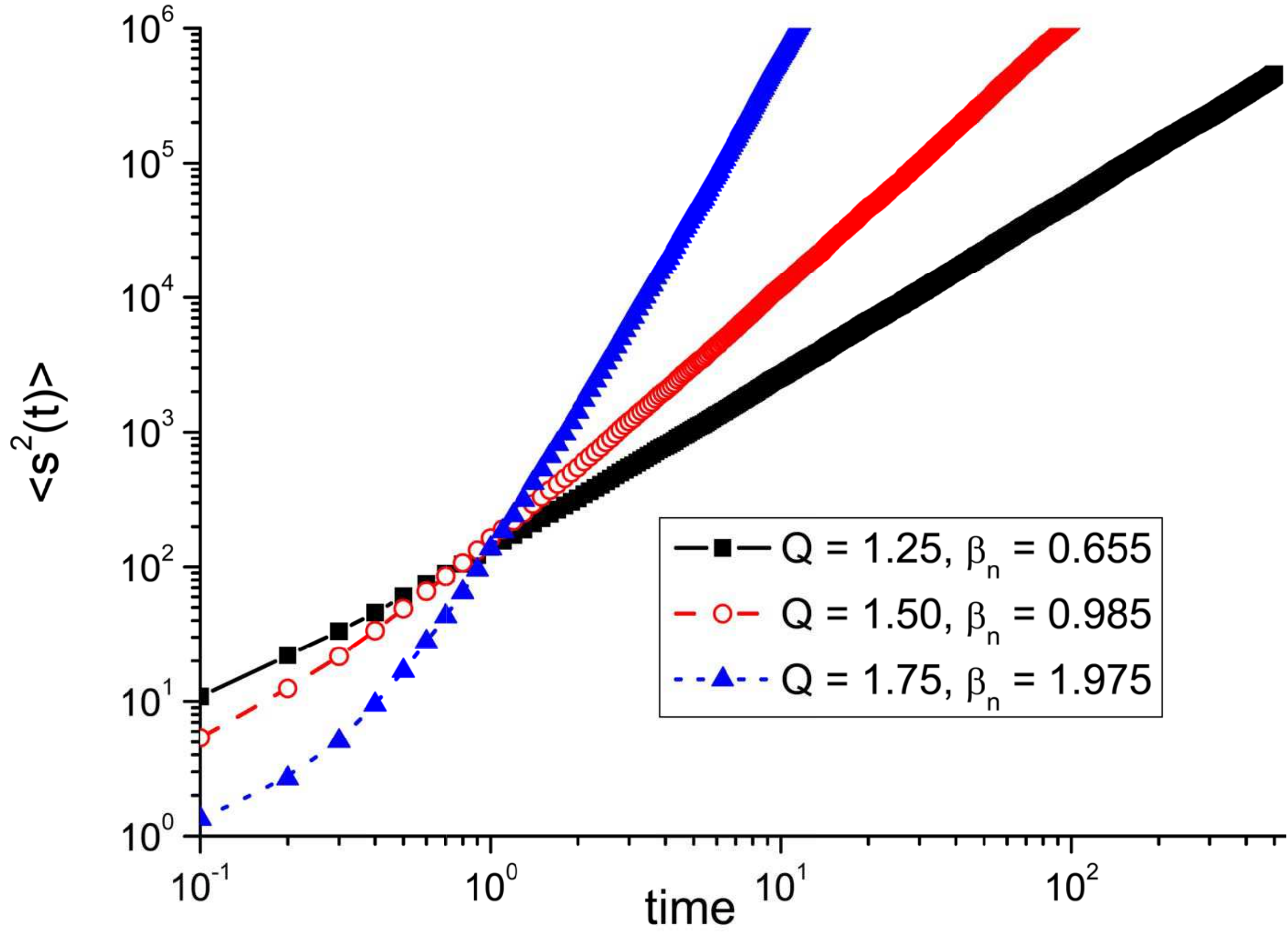}%
\hfill%
\includegraphics[width=0.46\textwidth]{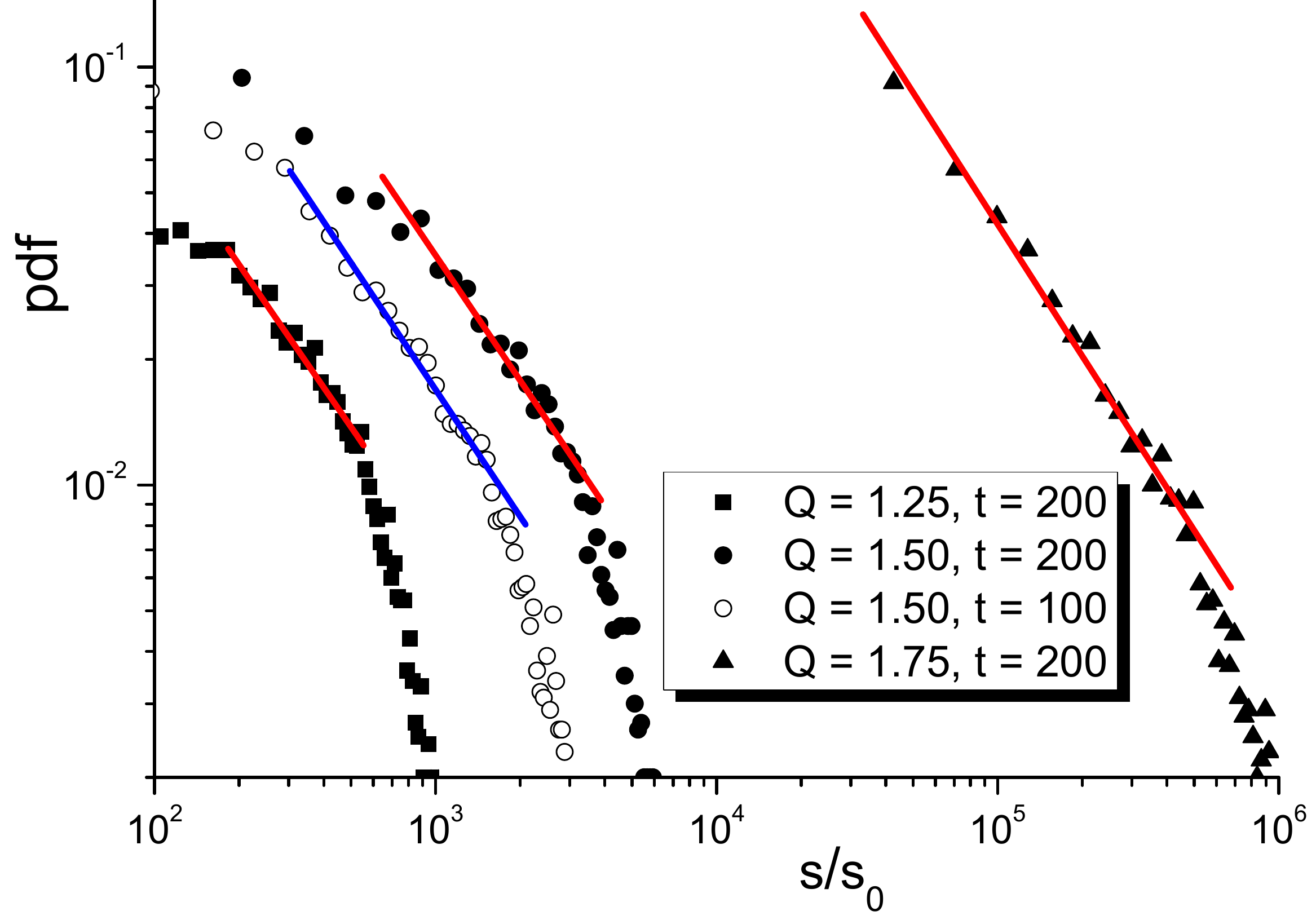}%
\\%
\parbox[t]{0.46\textwidth}{%
\centerline{(a)}%
}%
\hfill%
\parbox[t]{0.46\textwidth}{%
\centerline{(b)}%
}%
\vspace{-1mm}
\caption{(Color online) Protocol for the second moment
(a) and the corresponding pdfs (b) at different times (open and filled circles)
at fixed $Q$ and at fixed $t=200$ and different Tsallis exponent $Q$.
\label{num}}
\end{figure}

To describe the obtained universal dynamics and the universal behavior of the
distribution function over islands area that satisfy the above area conservation
condition, we apply the formalism of NFPE to our system. Since the island area
grows diffusively (by attachment/deattachment interacting atoms), we
start with the nonlinear diffusion equation for $\mathcal{N}(s,t)$ written in
the form
\begin{equation}\label{ndifN}
\partial_t \mathcal{N}=\nabla_{\mathrm{s}} \left[\mathcal{D}(\mathcal{N})\nabla_{\mathrm{s}}
\mathcal{N}\right],
\end{equation}
where we assume the form
$\mathcal{D}(\mathcal{N})=D_0\mathcal{N}^{1-Q}$; $D_0=\mathrm{const}$, $\nabla_{\mathrm{s}}\equiv
\partial/\partial s$ for generalized diffusivity. To obtain the corresponding time dependent solution, we
consider the system in an automodel regime assuming: $s(t)=a(t)y$,
$\mathcal{N}(s,t)=a^\varrho\varphi(y)$, where $a(t)$ is the scaling function
that measures the diffusion package smearing. Using the normalization condition and
the area conservation law, we immediately get $\varrho=-2$. Substituting such
constructions into equation~(\ref{ndifN}), we can separate a time dependent part and a
part depending on $y$:
\begin{equation}
 \dot a a^{1+2(1-Q)}=\lambda_0;\quad
  -\lambda_0\varphi=[\lambda_0 y \varphi + \varphi^{1-Q}\varphi']'.
\end{equation}
Here, $\lambda_0$ is the constant related to the time dependence. From the first
equation we get a relation between Tsallis parameter $Q$ and the scaling exponent
$\beta$ in the form
\begin{equation}\label{alpha}
\beta=[2(2-Q)]^{-1}.
\end{equation}
To solve the second equation, we assume a solution in the Tsallis form, i.e.,
$\varphi(y)=[C_0-\frac{1-Q}{D_0\delta}y^{\delta}]^{1/(1-Q)}$, where $C_0$,
$\delta$ will be defined. Next, using the relation $\varphi'=-D_0^{-1}
y^{\delta-1}\varphi^Q$, inserting it into equation for $\varphi$ we arrive at
the algebraic equation having a solution in the form
\begin{equation}
\varphi(y)=\left[\frac{\frac{y^{\delta}}{D_0}\left(1-\frac{y^{\delta-2}}{D_0\lambda_0}
\right)}{2-\frac{(\delta-1)}{D_0\lambda_0}y^{\delta-2}}\right]^\frac{1}{1-Q}.
\end{equation}
In further consideration, the quantity $1/D_0\lambda_0$ can be considered as a
small parameter of the theory. It allows us to obtain a reduced pdf of the
form $\varphi\propto y^{-1}$ with $C_0=0$ and $\delta=Q-1$. In such a case, the
desired pdf is
\begin{equation}\label{Nst}
\mathcal{N}(s,t)\approx
\left(\frac{t_0}{t}\right)^\frac{1}{2(2-Q)}\frac{s_0}{s}\,,
\end{equation}
where $t_0$ and $s_0$ depend on $D_0$ and relate to a normalization condition. It
is seen that depending on the parameter $Q$, one gets different dynamics for
islands area growth whereas the corresponding area distribution remains
universal, independently of the Tsallis parameter $Q$. According to the
obtained numerical data [see figures~\ref{NS}~(b), (c)] and equation~(\ref{alpha}) for the
Tsallis parameter, one has $1\leqslant Q < 2$. For the normal islands growth
characterized by $\beta=1$, we have $Q=3/2$. Therefore, the delayed dynamics
observed at an elevated deposition rate $F$ and at interaction energy $\varepsilon$
is characterized by $Q\in[1,3/2)$, whereas the enhanced dynamics observed at low
$F$ or $\varepsilon$ relates to $Q\in(3/2,2)$.
Using the above formalism for nonlinear diffusion equation for
$\mathcal{N}(s,t)$ we can write down the Langevin equation corresponding to
equation~(\ref{ndifN}) following \cite{LBorland,KK2005}. It takes the form
\begin{equation}\label{len}
\frac{{\rm d}s}{{\rm
d}t}=\sqrt{\mathcal{D}[\mathcal{N}(s,t)]}\xi(t)\equiv\left[t^{\frac{1}{2(2-Q)}}{s(t)}\right]^{\frac{Q-1}2}\xi(t),
\end{equation}
where $\xi(t)$ is the white noise having standard properties. Using its
formal solution together with the correlator
$\langle\xi(t)\xi(t')\rangle=\delta(t-t')$ in the automodel regime, we obtain
$\langle[s(t)-s(0)]^2\rangle\propto t^{2\beta(Q-1)+1}$. Comparing this time
dependence with \emph{a priori} known $\langle[s(t)-s(0)]^2\rangle\propto
t^{2\beta}$, one immediately arrives at the relation (\ref{alpha}). Testing
numerical solutions of the Langevin equation (\ref{len}) we have found a good
agreement between the analytical results and numerical data for time dependencies
of $\langle s^2(t)\rangle\propto t^{\beta_n}$, where $\beta_n$ is the fitting
exponent, and the corresponding pdfs with hyperbolic approximation $s^{-1}$ at
fixed times (see figure~\ref{num}). In our simulations, we took fixed $Q$ and
obtained $\beta(Q)\simeq \beta_n(Q)$ with errors up to 5\% [see protocols in
figure~\ref{num}~(a)]. The corresponding data for pdfs shown in figure~\ref{num}~(b) allow
one to elucidate that there is a domain for $s$ where universal behavior of pdf
is realized at different times and 
at different exponent $Q$. It follows that with an increase in $Q$ one gets
a growing interval for $s/s_0$ where universal $Q$-independent behavior of pdf is
realized.
\begin{figure}[!h]
\includegraphics[width=0.46\textwidth]{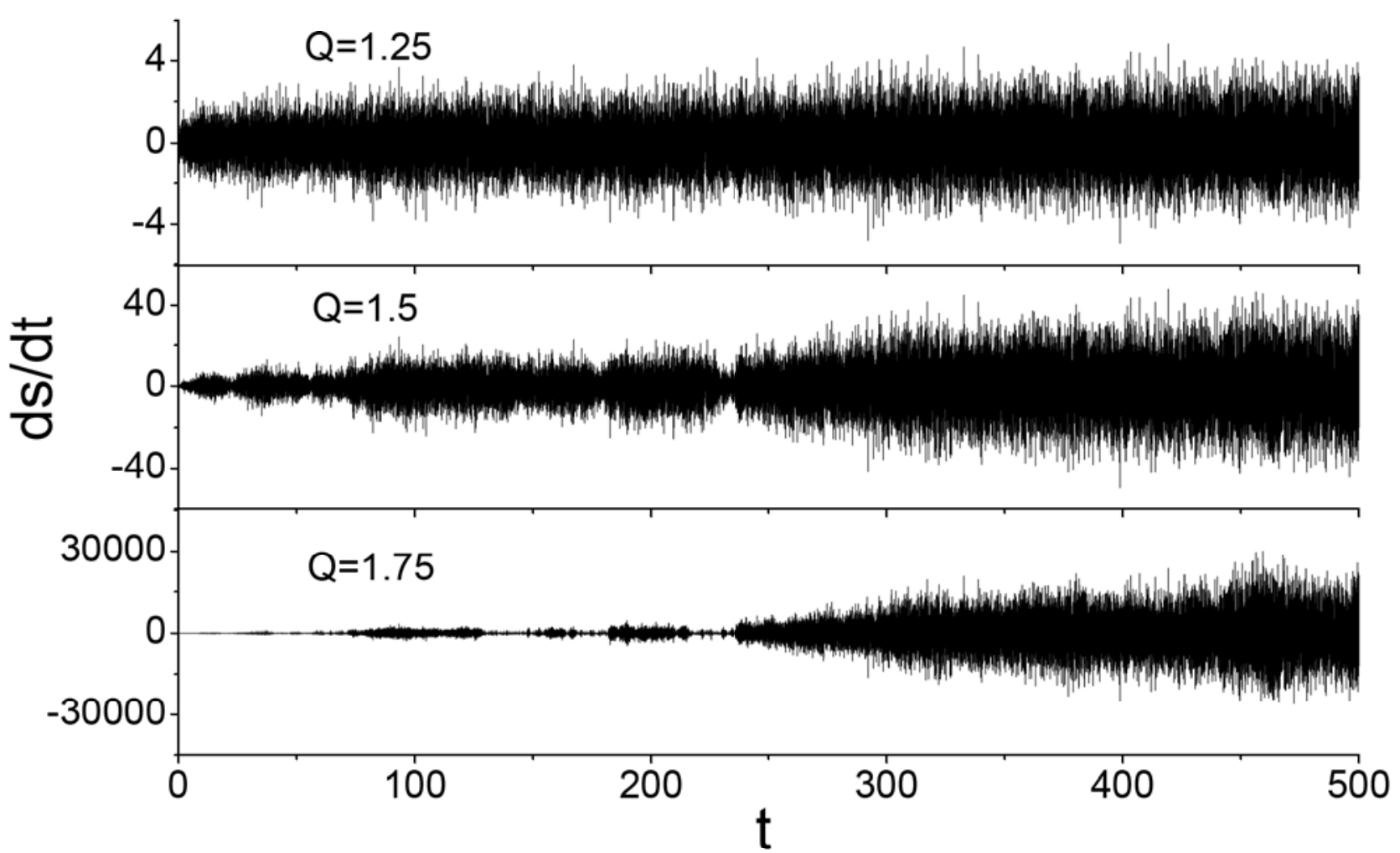}%
\hfill%
\includegraphics[width=0.46\textwidth]{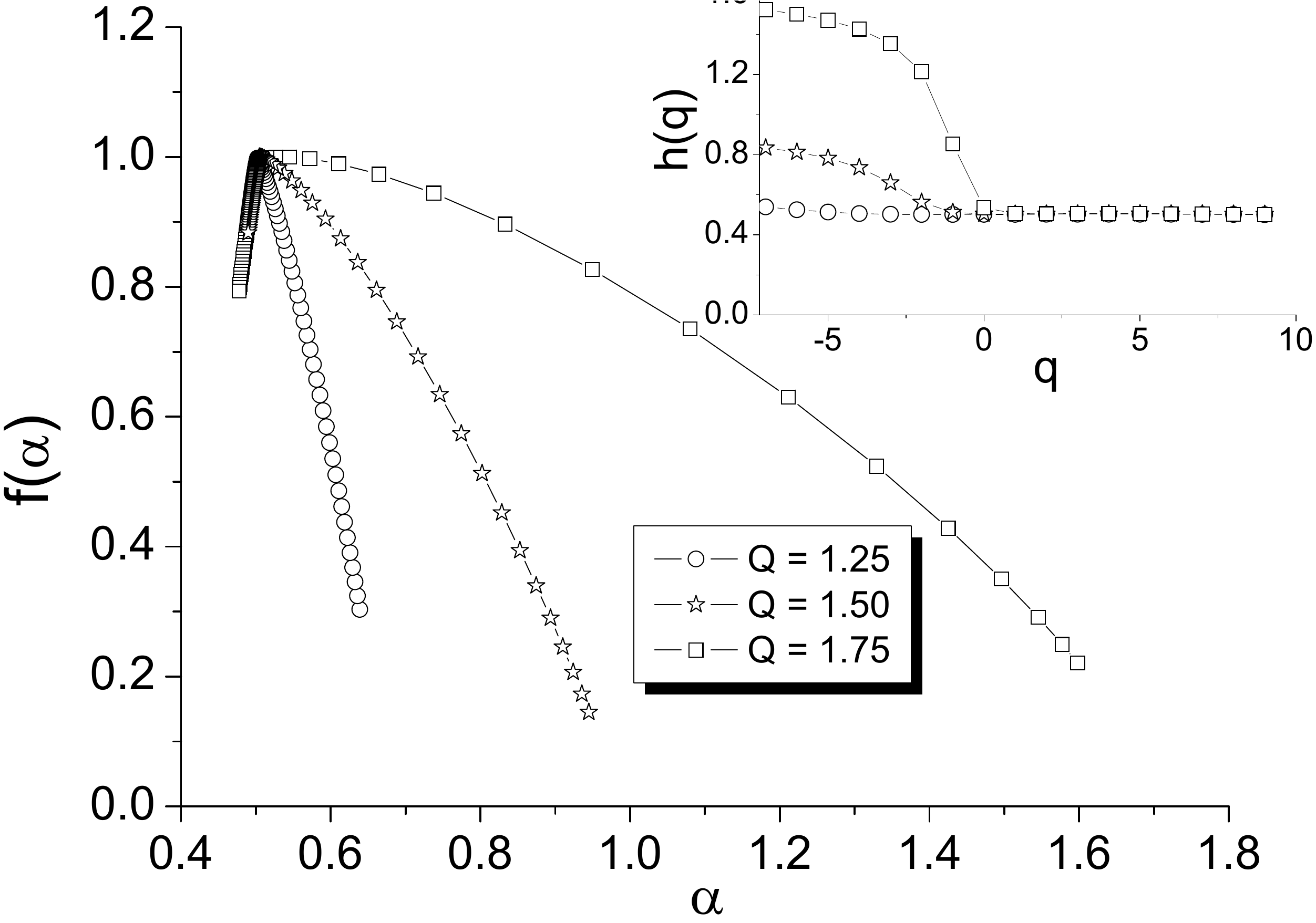}%
\\%
\parbox[t]{0.46\textwidth}{%
\centerline{(a)}%
}%
\hfill%
\parbox[t]{0.46\textwidth}{%
\centerline{(b)}%
}
\vspace{-1mm}
\caption{(a) Protocols for the
fluctuations in time series $s(t)$ at different $Q$ and (b) the singularity
spectrum $f(\alpha)$ with the dependencies $h(q)$ as insertion \label{ts}.}
\end{figure}

The jagging of time series $s(t)$ can be studied using multifractal detrended
fluctuation analysis (see \cite{KZKHBS}) as a generalization of the
standard multifractal theory \cite{Feder}. Following \cite{KZKHBS} we can
obtain a set of fractal exponents $h(q)$ and a singularity spectrum
$f(\alpha)=q[\alpha-h(q)]+1$ for multifractals, where $\alpha=h(q)+ qh'(q)$ is
the singularity strength, $q$ is the index for multifractality. It is known
that the quantity $h(q=2)$ coincides with the Hurst exponent $0\leqslant H\leqslant 1$
measuring the smoothness of the time series \cite{Hurst} and defining the fractal
dimension of time series as $D_{\mathrm{f}}=2-H$ \cite{Feder}. The corresponding protocols
of fluctuations ${\rm d}s(t)/{\rm d}t$ at different exponent $Q$ used in
detrended fluctuation analysis are shown in figure~\ref{ts}~(a). The related
dependencies $f(\alpha)$ and $h(q)$ are shown in figure~\ref{ts}~(b). It follows that
at $Q\gtrsim 1$, the corresponding fluctuations are characterized by small
variations in $h(q)\simeq 1/2$ and by narrow spectrum $f(\alpha)$ which means a weak
multifractality of the time series. In other words, at $Q\gtrsim 1$ one gets
Gaussianly distributed fluctuations. At elevated $Q$, the time series manifests
multifractal properties with wide spectrum $f(\alpha)$. The physical reason for
multifractal properties emergence lies in time correlations of the time series.
This result comes from the analysis of shuffled time series leading to the fact
that $h(q)$ for this series remains constant, 1/2. \cite{KZKHBS}.

\vspace{-3mm}

\section{Conclusions\label{s4}}

Using a nonlinear kinetic approach we have described universality and scaling
properties of pyramidal islands formation at epitaxial growth. In the framework
of the phase-field model for pyramidal islands growth in systems with
interacting adsorbate, we have analyzed the dynamics of the number of islands, their
size and the corresponding distribution functions. We proposed a generalized
theoretical model for the island size dynamics manifesting a self-similar behavior
and universality of the probability density over the island size. The corresponding
dynamics is studied using multifractal time series analysis. It is shown that
multifractality of the corresponding time series is related to time
correlations.

\vspace{-1cm}
\ukrainianpart

\title{Універсальність та самоподібна поведінка нерівноважних складних систем із нефіківською дифузією}%
\author{Д.О. Харченко, В.О. Харченко, С.В.Кохан}
\address{Інститут прикладної фізики НАН України,  вул. Петропавлівська 58, 40000 Суми, Україна}
\makeukrtitle
\begin{abstract}
\tolerance=3000%
Наведено аналітичні методи щодо опису нефіківської дифузії у нерівноважних
складних системах. Ці підходи застосовано до вивчення статистичних властивостей
росту пірамідальних острівків із взаємодіючим адсорбатом при епітаксіальному
рості. Використовуючи узагальнений кінетичний підхід, розглянуто
універсальність, скейлінгову динаміку та фрактальні властивості процесу росту
пірамід. У рамках узагальненої кінетики запропоновано теоретичну модель для
пояснення чисельно отриманих даних для середнього розміру островів, їх числа та
відповідної функції розподілу за розмірами.

\keywords
 складні системи, нелінійна дифузія, формування структур, фрактали
\end{abstract}

\end{document}